\documentclass[preprints,article,accept,moreauthors,pdftex]{Definitions/mdpi}  

\firstpage{1} 
\pubvolume{00}
\issuenum{1}
\articlenumber{5}
\pubyear{2020}
\copyrightyear{2020}
\history{Received: 24 September 2020; Accepted: 14 October 2020; Published: date}
\updates{yes} 

\Title{TITUS: Visualization of Neutrino Events in Liquid Argon Time Projection Chambers}

\Author{Corey Adams 
 $^{1,}$*\orcidA{} and Marco del Tutto $^{2}$\orcidB{}}

\AuthorNames{Corey Adams and Marco Del Tutto}

\address{%
$^{1}$ \quad Argonne National Labpratory, Lemont, IL, 60439 
 USA  \\
$^{2}$ \quad Fermi National Accelerator Laboratory, Batavia, IL, 60510 USA; mdeltutt@fnal.gov}

  \corres{Correspondence: corey.adams@anl.gov}

\abstract{The amount and complexity of data recorded by high energy physics experiments are rapidly growing, and with these grow the difficulties in visualizing such data. To study the physics of neutrinos, a type of elementary particle, scientists use liquid argon time projection chamber (LArTPC) detectors, among other technologies.  LArTPCs have a very high spatial resolution and resolve many of the elementary particles that come out of a neutrino interacting within the argon in the detector. Visualizing these neutrino interactions is of fundamental importance to understanding the properties of neutrinos, but also monitoring and checking on the detector conditions and operations. From these ideas, we have developed TITUS, an event display that shows images recorded by these neutrino detectors. TITUS is a piece of software that reads data coming from LArTPC detectors (as well as the corresponding simulation) and allows users to explore such data in multiple ways. TITUS is flexible to enable fast prototyping and customization.}


\graphicspath{{figures/}{pictures/}{images/}{./}}

\begin{document}


\section{Introduction}
Neutrinos are elementary particles that have many properties that are not yet understood. Many physics experiments are currently focused on studying neutrinos, and many of the current experiments are now employing a piece of detector technology called the liquid argon time projection chamber (LArTPC). These detectors are used to create digital images of the neutrino interactions with matter, which allow scientists to study and understand many of the properties of the neutrinos. 

\begin{figure}[H]
 \centering
 \includegraphics[width=0.7\textwidth]{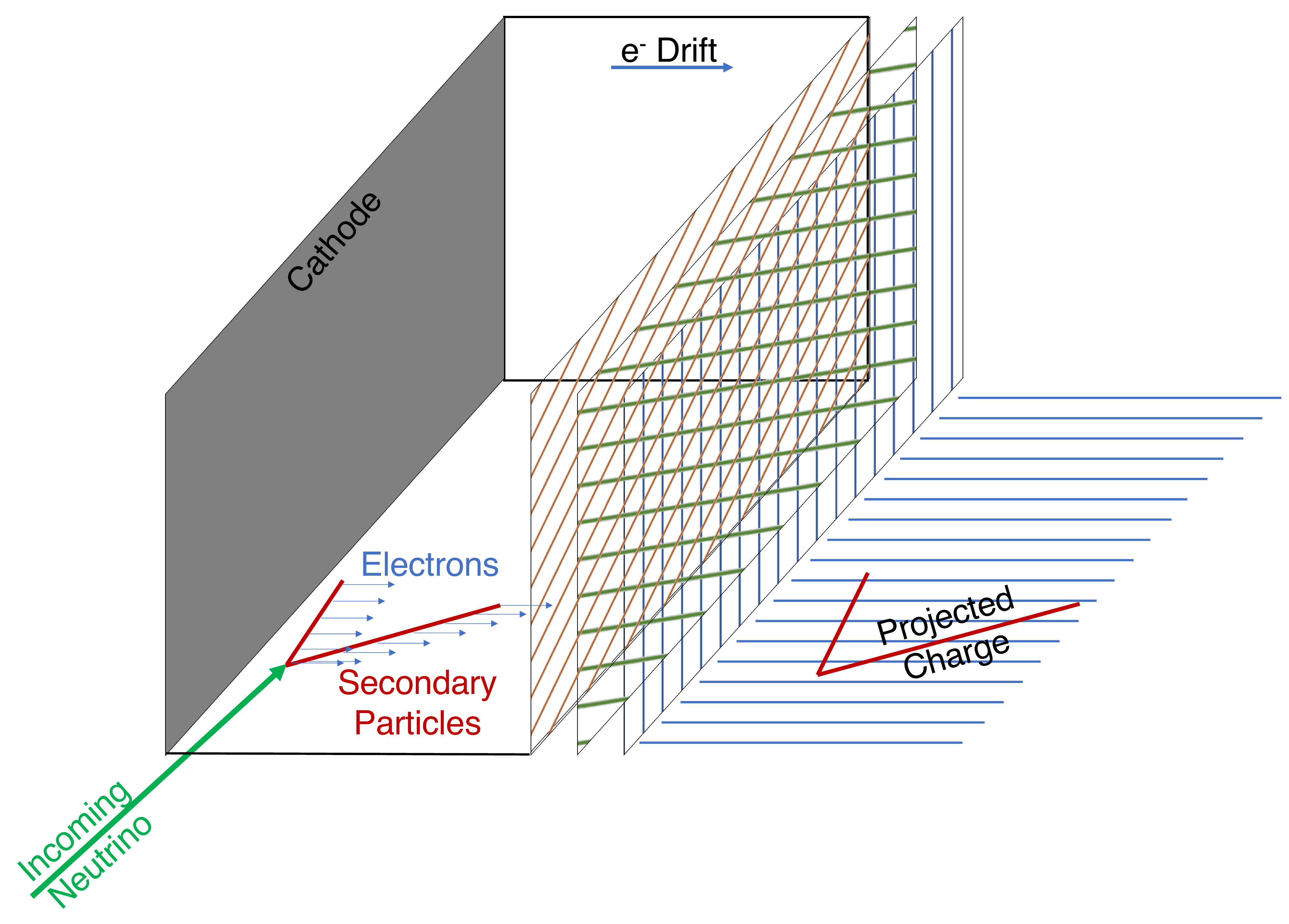}
 \caption{Operational principle of a generic LArTPC
.}
 \label{fig:lartpc}
\end{figure}

LArTPC detectors are at the forefront of experimental neutrino research, building larger and larger experiments, such as the Short Baseline Neutrino Program at Fermilab \cite{sbn}, the ProtoDUNE experiments at CERN \cite{protodune}, and eventually, the Deep Underground Neutrino Experiment (DUNE) \cite{dune_tdr_1, dune_tdr_2, dune_tdr_3}.  
The operational principle of a LArTPC detector is illustrated in Figure~\ref{fig:lartpc}. In LArTPC detectors, charged particles traversing the volume filled with liquid argon release trails of ionization electrons, which are transported to one side of the detector thanks to an electric field. Here, the electrons are read out thanks to a series of sense wires. There are usually two or more planes of sense wires and the electrons induce a signal on the first planes (orange and green in the picture) and are collected on the last plane (blue) where electric field lines terminate. The result is a collection of waveforms from every single wire. The~amplitude of such waveforms depends on the amount of detected electron charge, which changes as some particles are more ionizing than others. 
Placing all the waveforms next to each other gives rise to an image, and in the end, each LArTPC digitizes a collection of 2D image-like signals, across several projections of 3D space corresponding to the wire orientation on the different wire planes (in~the figure, the images result in a top-down projection, and at two angles offset from vertical). The~waveforms that make the images are digitized at up to 2~MHz timing resolution. The collection of projected images from the same 3D interaction is referred to as an {event}.
In the smallest detector of those mentioned above, the images produced have a resolution of ($2400 \times 9600$) pixels for two images at angles, and a resolution of ($3456 \times 9600$) pixels for the vertical image.  

With these detectors that provide such incredible insights into the neutrinos, a visualization tool is required to display the images of the neutrino interactions.  The tool must be interactive and accommodate the unique nature of this data (multiple simultaneous projections of 3D space), as well as the large image size.  Additionally, for the analysis of this data, we must also visualize the higher-level data products derived from these images, such as the detected charge from particle interactions, the~clustered depositions, and even the projection of objects from 3D (from high-level analysis) onto the 2D planes.

\begin{figure}[H]
 \centering
 \includegraphics[width=0.98\textwidth]{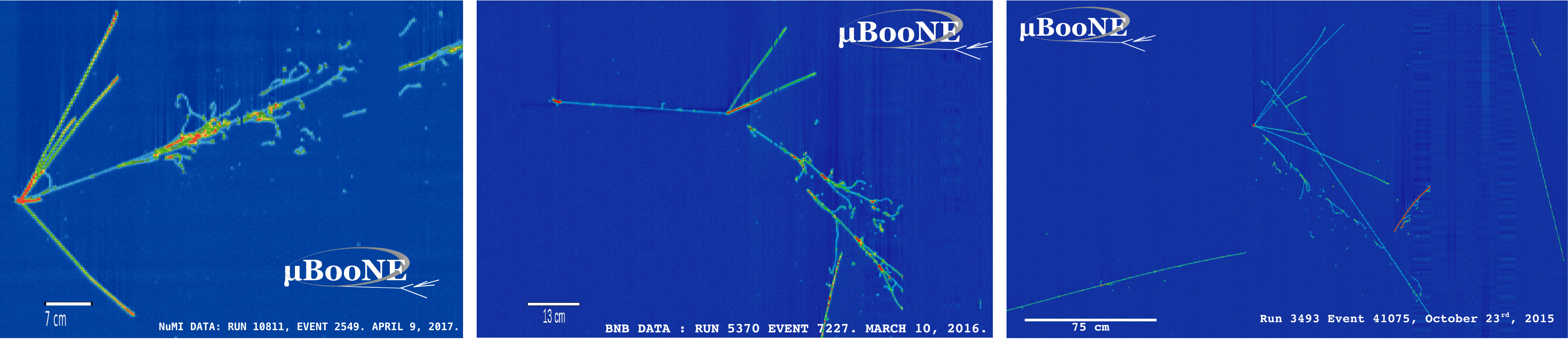}
 \caption{TITUS displays of neutrino candidate interactions with matter as recorded by the MicroBooNE experiment~\cite{ub_plots}.}
 \label{fig:teaser}
\end{figure}

From these ideas comes TITUS: an event display to visualize neutrino interactions in time projection chambers. This display is currently in use in many neutrino experiments at the Fermi National Accelerator Laboratory: MicroBooNE, SBND, and ICARUS~\cite{uboone, sbn, icarus}.  Additionally, techniques developed for TITUS have been extended to visualize other neutrino experiments, including ProtoDUNE and LArIAT \cite{lariat}. Three examples of event visualization with TITUS are shown in Figure~\ref{fig:teaser}.

The philosophy behind TITUS is to develop a software package that allows users to easily customize the displayed objects and quickly add new ones that may be needed by particular physics analyses. Each type of data is rendered with only a few lines of \texttt{Python}, meaning a user can easily understand and extend a rendering routine.  As an example, it is trivial to change the color of reconstructed tracks if they are reconstructed as "neutrino" in origin instead of  "cosmic" background.  The~speed and simple layout of this viewer, alongside  its simple customization, significantly shortens the time it takes to go from a raw data file to visualized content, hence speeding up physics~analyses.

\section{Related Work}

LArTPC technology has been in large scale experiments, beyond the R\&D phase, for fundamental physics for nearly two decades, and we are not the first to visualize the data.  We took inspiration from the visualization tools of \texttt{larsoft} \cite{larsoft} and ICARUS's QScan \cite{qscan}, but recreated from scratch a modern tool that incorporates successes from the past with modern software practices.

The visualization tool from the \texttt{larsoft} software package allowed users to accurately image neutrino interactions; however, it contained fundamental shortcomings.  First, images are based on rendering 2D histograms from the ROOT \cite{root} package, which was slow and memory intensive.  Second, the configuration of the viewer was C++ based and not trivial to modify or customize by a user.

A predecessor of the \texttt{larsoft} event display is \texttt{QScan}~\cite{qscan}, a tool developed by the ICARUS collaboration.
\texttt{QScan} used \texttt{Qt} \cite{qt} to render images and create interactions with user, a practice we have adopted with \texttt{TITUS} as well. However, \texttt{QScan} was developed exclusively for the ICARUS collaboration, and we have developed instead a cross-platform tool that works with multiple experiments in neutrino physics.  Further, \texttt{QScan} is a pure C++ program, but we have implemented a \texttt{Python}-based front-end to lower the customization barrier for users.

\section{Challenges}


There are several unique challenges to developing a tool to accurately and efficiently display a LArTPC image set, while most importantly enhancing the ability of a physicist to understand and analyze the data.  First, a physicist is typically interested in viewing a single event in an intuitive way, including all aspects of that event.  This includes raw data, which are typically large and cumbersome for traditional rendering programs to process.  This also includes derived data products, including, for example, reconstructed locations of signal depositions or the projection of a reconstructed 3D track onto original images for validation.  LArTPC detectors have a non-trivial geometry, and the complexities of producing a mapping from 2D to 3D in a visual tool encourage a dedicated tool such as TITUS.

Additionally, the large collaborative structure of these experiments implies that one physicist may not necessarily be aware of all of the derived data products, nor aware of the relations between them.  This challenge naturally does not apply only to neutrino physics, but to any complex dataset.  Nevertheless, in TITUS we have striven to present an intuitive, easy-to-use application that presumes no prior knowledge from the reader.  Anything available to be visualized is easy to do so, with~nothing~"hidden."

A last obstacle for a user to visualize neutrino data is the data format itself.  While there are very many visualization techniques available in nearly every programming language, most assume the user is able to easily retrieve the raw objects from file that they want to visualize.  In the case of neutrino physics, the large and complex datasets have given rise to dedicated data structures that are unique, and consequently, not intuitive to a non-expert.  The field of visualization, applied to neutrino physics, therefore, requires tools that lower the barrier to entry for physicists in all aspects: geometrical complexity, data availability, and data access.  TITUS specifically addresses all of these aspects.

\section{Data Description}

Data are stored in \texttt{ROOT} \cite{root} files with custom layers produced by \texttt{larsoft}, built upon the \texttt{art} framework \cite{art}.  Access of the data is performed using \texttt{gallery} \cite{gallery}, a lightweight access tool for \texttt{art} data products.  There are no images stored in the files, but rather, each column of pixels is stored as an array, with columns in arbitrary order.  This arises from the data collection and digitization scheme, which can be explored further here \cite{uboone_signal_processing_1, uboone_signal_processing_2}.  Each data file contains several events, and each event is a collection of raw image data along with derived objects and metadata. Metadata includes high-level event indexing, typically denoted with "run," "subrun," and "event" indexes, as well as timestamps for data collection and whether the event is recorded from a detector or simulated.

Derived data products, such as the identification of charge depositions in a column of pixels (referred to as a "hit") are typically created in offline reconstruction and analysis modules in the \texttt{larsoft} package (such as the GaussHitFinder \cite{Baller_2017}).  Products are stored via a standardized serialization model, which for "hits" includes the central location, width, peak, and corresponding uncertainties.  While the algorithm to find charge depositions (or other data products) may vary, the data product model is fixed by the \texttt{larsoft} package.  The raw image data, stored in a column of pixels referred to as a "wire" (the name comes from the wire-based readout system in the physical detectors) are connected to the derived "hits" through associations of data products in \texttt{larsoft}.

\section{System Architecture and Implementation}

The TITUS display is designed to be fast, intuitive, and extensible.  To support this, we have developed the entire interface through \texttt{PyQt} \cite{pyqt,qt}, and use the \texttt{PyQtGraph} \cite{pyqtgraph} package to efficiently render images with OpenGL.  The \texttt{larsoft} package is exclusively C++, while we want to enable customization and extensibility through \texttt{Python}.  To do this, we have created C++ processing kernels, along with wrappers to enable them in \texttt{Python}, enabling optimized data fetching with easy access to the data products in \texttt{Python}.  For each event to be visualized, the data processing kernel is enabled by the GUI through the actions of the user: if a user selects raw images or derived products to be drawn (typically through a drop-down menu in the GUI), the appropriate kernel is added to the list of processing kernels and it's visualization routines are called.

At each change of event, or when the user changes the features to be drawn, the event is re-processed and re-rendered.  The user also has the ability to jump to random events in the file they are using, allowing the physicist user to seek events of interest.  When data products are present in a file but not requested to be drawn, they are not read from disk, which enables fast loading of the viewer and similarly fast random access of events from files.

\section{Data Processing}

\begin{figure}[H]
 \centering
 \includegraphics[width=0.98\textwidth]{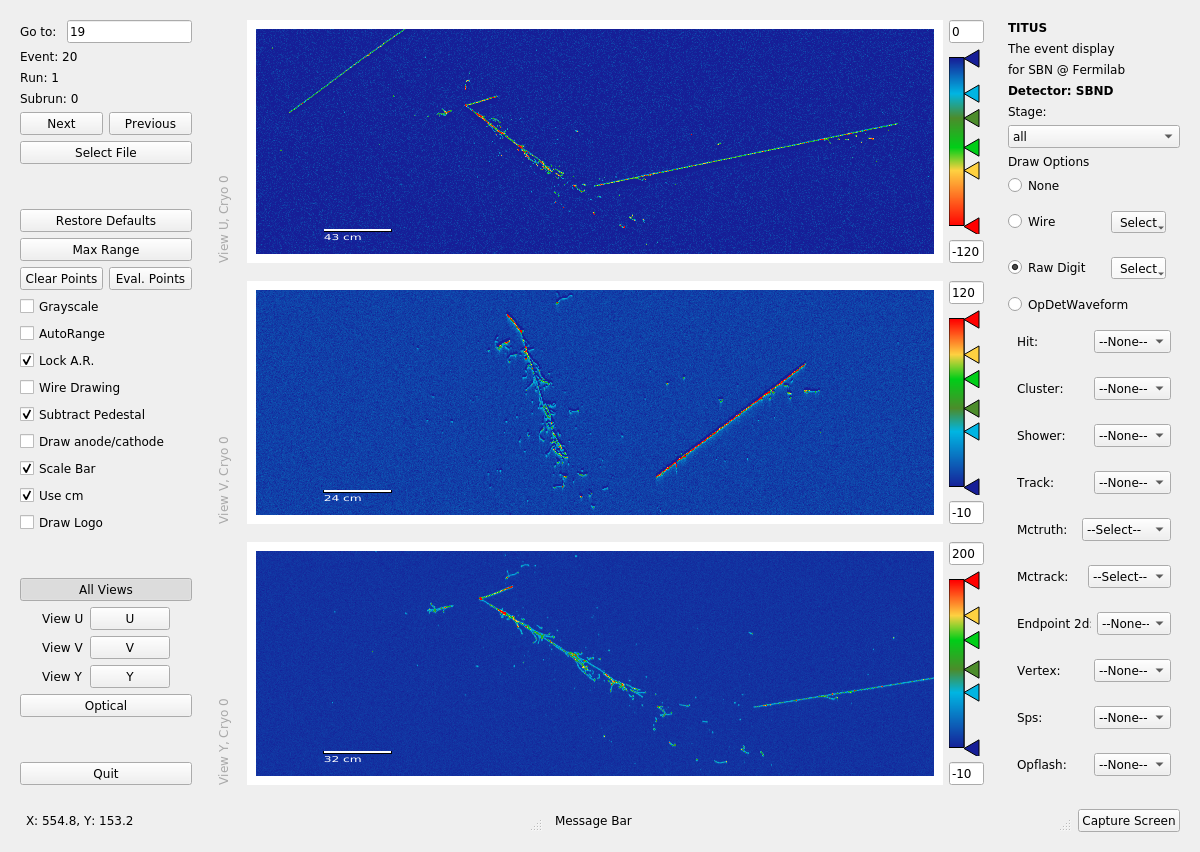}
 \caption{The main view of the TITUS event display. Here showing a simulated neutrino event in the SBND detector~\cite{sbn}.}
 \label{fig:evd_image}
\end{figure}

Upon the first opening of a file, TITUS scans all available data products in the file and creates a list of identifiers for each data product.  As experimental software, with diverse research teams designing multiple algorithms to produce similar derived data products, the \texttt{larsoft} framework supports creation and serialization of multiple instances of each data product.  As an example, there are several algorithms to find the depositions of charge upon each sense wire, and therefore multiple independent collections of "hits."  TITUS will identify all instances of "hits" data products (for example), and collect their identification string for the user to select from.  In the right column of the display, seen in Figure~\ref{fig:evd_image}, the user will see "- -None- -" for data products not present in the file, and "- -Select- -" for data products that are available in the file.

For each data-processing kernel the user has selected, a minimal amount of data processing is performed.  Typically, the data is retrieved from file as an {\verb std::vector<[DataProduct]> } object, where the data product contains significantly more information than needs to be rendered.  The TITUS processing kernels, therefore, inspect the data product arrays, copy only the necessary information to a structure, and expose that structure to \texttt{Python} to render via the PyQt package.

\begin{figure}[H]
 \centering
 \includegraphics[width=0.7\textwidth]{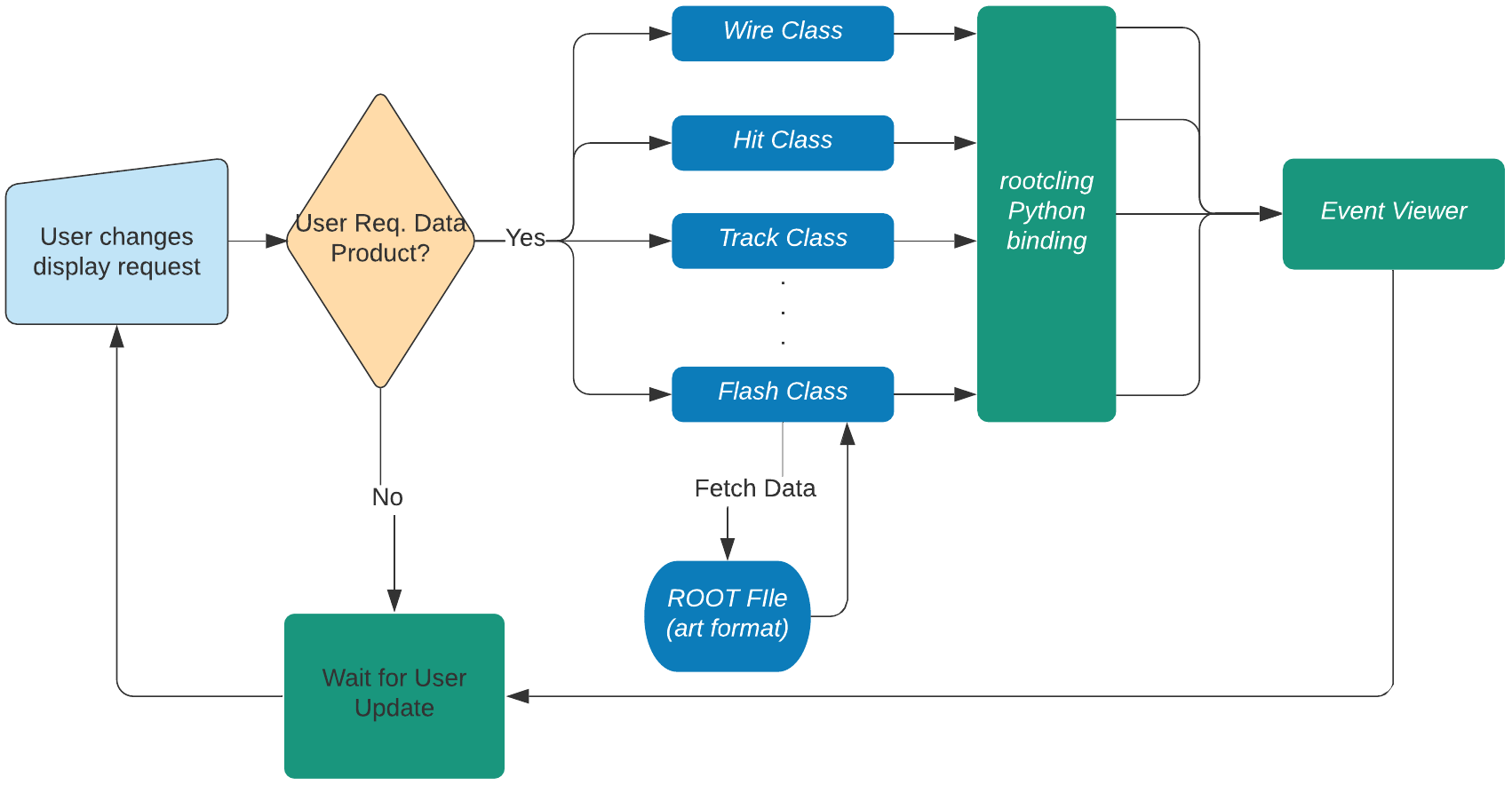}
 \caption{Data processing. The blue part is written in C++, while the green part is in \texttt{Python}.}
 \label{fig:flowchart}
\end{figure}

Exceptions to the processing scheme are the image-like data products, known as "rawdigit" for the raw detector data, and "wire" for the de-convolved and noise-removed image data.  In this case, TITUS reads each wire (which do not necessarily come in order) and copies the raw data into a memory buffer large enough to hold the entire image.  The buffer is then wrapped in a NumPy { \verb PyArray_Object }, which is passed to the \texttt{Python} rendering tools. A flowchart of the data processing is shown in Figure~\ref{fig:flowchart}.

\section{Visual Design}

The main user interface is encapsulated in several \texttt{Python} classes to build a centralized view window with surrounding interactive features.  As can be seen in Figure~\ref{fig:evd_image}, the bulk of the display is used for visualizing the data, with the ability to pan, zoom, stretch, and export the rendered images.  The~user can visualize the event using the same spatial scale of different scales independently, one per projection.  All data products selected to be rendered are viewed in the central display and derived data types (like "hits") can be overlaid on the images.  As the user mouses over the display, the~information bar at the bottom updates with the location (in detector coordinates) of the mouse in the image. The~central view area uses a rainbow based heat map for visualizing the data, which is common in this type of application, however it fully customizable by modifying the appropriate \texttt{Python} class. 

Additionally, several derived data products will display their metadata when a "mouse hover" action is used.  Surrounding the central view area is a suite of controls to enable full use of the display.

\textbf{Event Control:} 
Users must be able to easily select an event within a file for random access.  In the top left of the display, we allow users to toggle through a file in forward or reverse order or jump, at random, to an event.  Users may also easily select a new file from here.

\textbf{Drawing Options:}
In the central left location, we have collected all of the most common whole-image operations, including but not limited to pedestal subtraction for raw data, inclusion of a scale bar, and the ability to add additional auxiliary information (logo, timestamps) for producing publication-quality images.  

\textbf{Plane Control:}
 Each event, as mentioned, includes several projected 2D images of a 3D space.  Here, we allow users to disable some views interactively to render other projections more clearly.  Additionally, these~technologies include an auxiliary detection system, called the "optical" system, which can be viewed using this control area for the same event.

\textbf{Data Product Rendering Control:}
As mentioned above, there are multiple derived data types from the raw image-like data.  On the right side of the display, the user is given full control over what data products are rendered in the view.  At the top is a filter for the data processing "stage", since multiple instances of a single data product can be produced during the data processing.  The full enumeration of derived data products is beyond the scope of this paper, however, all "official" \texttt{larsoft} data products are supported by TITUS if they represent objects that may be visualized.


\section{Case Studies}

There are two main scenarios in which this event display is used: to constantly monitor the raw data coming from a LArTPC detector and to analyze the data afterward.

TITUS visualizes the MicroBooNE detector's data in a constant stream, 24/7, and it shows the latest data recorded. This enables operators in the experiment's control room to make sure that the raw data have an acceptable quality, and allows them to immediately identify potential issues with the detector. The event display shows stability over long periods of time and it is fast and intuitive to use for control-room operators who often have to act quickly to understand and fix potential issues. SBND and ICARUS, as of this publication, are not yet taking data.

The MicroBooNE, SBND, and ICARUS collaborations are also using TITUS for the second, offline application. Here, TITUS is used when the data being analyzed have already been recorded, processed, and potentially (though not necessarily) reconstructed. Analysts can visualize more complex and derivative products, like particle tracks in the detector. TITUS has shown its success by being an important tool for many analyzers to understand the data and to bring physics analyses to completion.

\section{Performance}

The main performance considerations for TITUS are two-fold. First, the software must quickly load and display data from file with no significant latency, even for large image sizes.  Second, the~display must be smooth and responsive to pan/zoom and other interactive measures.

To address the latency of file loading, we have ensured fast performance with C++ based IO interactions, which are wrapped in \texttt{Python} for use in the display.  Each time TITUS needs to fetch data from file, only the data requested by the user are loaded to ensure the fastest loading time.  Particularly intensive data, such as raw images, are read directly into \texttt{numpy} \cite{numpy} arrays using the Python-C API.

Further, the GUI is built on PyQt and all rendered objects leverage the Qt framework for visualization.  In this way, we leverage industry-standard tools to ensure smooth and responsive performance.  The rendering of 1D plots, and fast render of 2D images, is simplified with the use of \texttt{pyqtgraph} \cite{pyqtgraph}.

\section{Conclusions}

In this paper, we have presented TITUS, a piece of software for visualization of data coming from LArTPC detectors. TITUS is a cross experiment software, and it is currently used by several collaborations. We have described the system architecture, the design, and the implementation, each showing how TITUS meets the challenging mission of providing a fast event viewer that can work across different experiments, and that allows users easy customization to tailor specific experiment's needs.

TITUS has been used extensively by the MicroBooNE collaboration, both in online and offline operations.  This tool has enabled quick review and checking of their detector data as it comes in.  Additionally, it has been used to validate and check the reconstructed data products produced by MircoBooNE's physics analyses.  Further, TITUS has developed an efficient and extensible software framework that allows it to visualize data from SBND and ICARUS experiments, though they have different geometries from MicroBooNE.  

Rendering leverages open source, standard tools such as \texttt{Qt} and \texttt{Python}.  The user interface, designed to be intuitive and flexible, was built via \texttt{PyQT} and \texttt{pyqtgraph}, leveraging the efficiency of \texttt{numpy} for data manipulation and \texttt{OpenGL} for 2D rendering.  In summary, TITUS delivers performance, ease of use, and portability across experiments, leading to its adoption in multiple LArTPC neutrino~experiments.

\vspace{6pt} 



\authorcontributions{Conceptualization, C.A.; Project administration, M.d.T.; Software, C.A. and M.d.T.; Visualization, C.A.; Writing – original draft, M.d.T. All authors have read and agreed to the published version of the manuscript.} 


\acknowledgments{We gratefully thank the SBND collaboration for giving us access to their simulations.
The~submitted manuscript has been created by UChicago Argonne, LLC, Operator of Argonne National Laboratory ("Argonne"). Argonne, a U.S. Department of Energy Office of Science laboratory, is operated under contract number DE-AC02-06CH11357.
Additionally, this document was prepared using the resources of the Fermi National Accelerator Laboratory (Fermilab), a U.S. Department of Energy, Office of Science, HEP User Facility. Fermilab is managed by Fermi Research Alliance, LLC (FRA), acting under contract number DE-AC02-07CH11359.}

\funding{This research received no external funding.}
\conflictsofinterest{The authors declare no conflict of interest.} 
\reftitle{References}


\begin{thebibliography}{999}
\providecommand{\natexlab}[1]{#1}

\bibitem[Acciarri \em{et~al.}(2015)Acciarri et~al.]{sbn}
Acciarri, R.; Adams, C.; An, R.; Andreopoulos, C.; Ankowski, A.M.; Antonello, M.; Asaadi, J.; Badgett, W.; Bagby, L.; Baibussinov, B.; et al.
\newblock A Proposal for a Three Detector Short-Baseline Neutrino Oscillation
  Program in the Fermilab Booster Neutrino Beam. \emph{arXiv}  \textbf{2015},
arXiv:1503.01520.

\bibitem[Abi \em{et~al.}(2017)Abi et~al.]{protodune}
Abi, B.; Acciarri, R.; Acero, M.A.; Adamowski, M.; Adams, C.; Adams, D.L.; Adamson, P.; Adinolfi, M.; Ahmad, Z.; Albright, C.H.; et al.
\newblock The Single-Phase {ProtoDUNE} Technical Design Report. \emph{arXiv} {\bf 2017}, arXiv:1706.07081.

\bibitem[Abi \em{et~al.}(2018{\natexlab{a}})Abi et~al.]{dune_tdr_1}
Abi, B.; Acciarri, R.; Acero, M.A.; Adamowski, M.; Adams, C.; Adams, D.; Adamson, P.; Adinolfi, M.; Ahmad,~Z.; Albright, C.H.; et al.
\newblock {The DUNE far detector interim design report, Volume 1: Physics,
  technology and strategies}. \emph{arXiv}  {\bf 2018}, arXiv:1807.10334.

\bibitem[Abi \em{et~al.}(2018{\natexlab{b}})Abi et~al.]{dune_tdr_2}
Abi, B.; Acciarri, R.; Acero, M.A.; Adamowski, M.; Adams, C.; Adams, D.; Adamson, P.; Adinolfi, M.; Ahmad, Z.; Albright, C.H.; et al. 
\newblock {The DUNE far detector interim design report, Volume 2: Single-phase
  module}. \emph{arXiv}  {\bf 2018}, arXiv:1807.10327.
 

\bibitem[Abi \em{et~al.}(2018{\natexlab{c}})Abi et~al.]{dune_tdr_3}
Abi, B.; Bansal, S.; Friedl, A.; Kocaman, B.; Djurcic, Z.; Goudzovski, E.; Rakotondravohitra, L.; Salukvadze,~G.; Mazzucato, E.; Densham, C.; et al.
\newblock {The DUNE far detector interim design report, Volume 3: Dual-phase
  module}. \emph{arXiv} {\bf 2018}, arXiv:1807.10340.
 

\bibitem[Acciarri \em{et~al.}(2017)Acciarri et~al.]{uboone}
Acciarri, R.; Adams, C.; An, R.; Aparicio, A.; Aponte, S.; Asaadi, J.; Auger, M.; Ayoub, N.; Bagby, L.; \mbox{Baller, B.; et al.}
\newblock {Design and Construction of the MicroBooNE Detector}.
\newblock {\em JINST} {\bf 2017}, {\em 12},~P02017.
  

\bibitem[Benetti \em{et~al.}(1993)Benetti et~al.]{icarus}
Benetti, P. 
\newblock A three-ton liquid argon time projection chamber.
\newblock {\em Nucl. Instruments Methods Phys. Res. Sect. A 
  Accel. Spectrometers Detect.  Assoc. Equip.} {\bf 1993},
  {\em 332},~395 -- 412.

\bibitem[Acciarri \em{et~al.}(2019)Acciarri et~al.]{lariat}
Acciarri, R.; Adams, C.; Asaadi, J.A.; Backfish, M.; Badgett, W.; Baller, B.; Rodrigues, O.B.; Blaszczyk, F.D.; Bouabid, R.; Bromberg, C.; et al.
\newblock {The Liquid Argon In A Testbeam (LArIAT) Experiment}. \emph{arXiv} {\bf 2019},  
 arXiv:1911.10379.

\bibitem[ub_()]{ub_plots}
{MicroBooNE} Website.
\newblock  Available online: \url{https://microboone.fnal.gov}  (accessed on 18 May 2020).
 

\bibitem[Snider and Petrillo(2017)]{larsoft}
Snider, E.L.; Petrillo, G.
\newblock {LArSoft: Toolkit for Simulation, Reconstruction and Analysis of
  Liquid Argon TPC Neutrino Detectors}.
\newblock {\em J. Phys. Conf. Ser.} {\bf 2017}, {\em 898},~042057.


\bibitem[Lussi(2011)]{qscan}
{Lussi, D.}
\newblock QScan. 2011. Available online: \url{https://indico.cern.ch/event/129268/contributions/1348759/attachments/88181/126234/Lussi_Qscan_GLA2011.pdf} (accessed on 11 July 2020).


\bibitem[Brun and Rademakers(1997)]{root}
Brun, R.; Rademakers, F.
\newblock {ROOT---An Object Oriented Data Analysis Framework}.
\newblock {\em Nucl. Inst.  Meth.  Phys. Res.~A} {\bf 1997}, {\em
  389},~81--86.

\bibitem[qt()]{qt}
Qt.
\newblock  Available online: \url{https://www.qt.io}  (accessed on 8 June 2020).
 
\bibitem[Green \em{et~al.}(2012)Green, Kowalkowski, Paterno, Fischler, Garren,
  and Lu]{art}
Green, C.; Kowalkowski, J.; Paterno, M.; Fischler, M.; Garren, L.; Lu, Q.
\newblock {The Art Framework}.
\newblock {\em J. Phys. Conf.~Ser.} {\bf 2012}, {\em 396},~022020.

\bibitem[gal()]{gallery}
{Gallery}.
\newblock  Available online: \url{https://art.fnal.gov/gallery/}  (accessed on 12 October 2020).
 

\bibitem[Adams \em{et~al.}(2018{\natexlab{a}})Adams
  et~al.]{uboone_signal_processing_1}
Adams, C.; An, R.; Anthony, J.; Asaadi, J.; Auger, M.; Balasubramanian, S.; Baller, B.; Barnes, C.; Barr, G.; Bass, M.; et al.
\newblock Ionization electron signal processing in single phase {LArTPCs}. Part
  {II}. Data/simulation comparison and performance in {MicroBooNE}.
\newblock {\em J.  Instrum.} {\bf 2018}, {\em
  13},~P07007.

\bibitem[Adams \em{et~al.}(2018{\natexlab{b}})Adams
  et~al.]{uboone_signal_processing_2}
Adams, C.; An, R.; Anthony, J.; Asaadi, J.; Auger, M.; Bagby, L.; Balasubramanian, S.; Baller, B.; Barnes, C.; Barr, G.; et al.
\newblock Ionization electron signal processing in single phase {LArTPCs}. Part
  I. Algorithm Description and quantitative evaluation with {MicroBooNE}
  simulation.
\newblock {\em J.  Instrum.} {\bf 2018}, {\em
  13},~P07006.

\bibitem[Baller(2017)]{Baller_2017}
Baller, B.
\newblock Liquid argon {TPC} signal formation, signal processing and
  reconstruction techniques.
\newblock {\em J. Instrum.} {\bf 2017}, {\em
  12},~P07010.

\bibitem[pyq({\natexlab{a}})]{pyqt}
{PyQt}.
\newblock Available online: \url{https://riverbankcomputing.com/software/pyqt/} (accessed on 8 June 2020).

\bibitem[pyq({\natexlab{b}})]{pyqtgraph}
{PyQtGraph}.
\newblock  Available online:  \url{http://www.pyqtgraph.org}  (accessed on 8 June 2020).
 

\bibitem[Harris \em{et~al.}(2020)Harris, Millman, van~der Walt, Gommers,
  Virtanen, Cournapeau, Wieser, Taylor, Berg, Smith, Kern, Picus, Hoyer, van
  Kerkwijk, Brett, Haldane, del R{\'\i}o, Wiebe, Peterson, G{\'e}rard-Marchant,
  Sheppard, Reddy, Weckesser, Abbasi, Gohlke, and Oliphant]{numpy}
Harris, C.R.; Millman, K.J.; van~der Walt, S.J.; Gommers, R.; Virtanen, P.;
  Cournapeau, D.; Wieser, E.; Taylor,~J.; Berg, S.; Smith, N.J.; et al.
\newblock Array programming with NumPy.
\newblock {\em Nature} {\bf 2020}, {\em 585},~357--362.

\end{thebibliography}
\end{document}